\documentstyle[multicol,prl,aps,graphics]{revtex}
\begin{document}
\title{Metal-insulator transitions in systems with  
electron-phonon and Coulomb interactions }

\author{J.E. Han and O. Gunnarsson}   
\address{Max-Planck-Institut f\"ur Festk\"orperforschung, 
D-70506 Stuttgart, Germany}

\date{\today}
\maketitle
\begin{abstract}
We consider a model which includes the electron-phonon coupling 
to A$_g$ or Jahn-Teller H$_g$ phonons, the Coulomb interaction $U$ and
an exchange integral $K$. We study the metal-insulator transition
for the integer fillings $n=3$ and 4. We find that the coupling 
to A$_g$ phonons increases the critical value $U_c$ where the transition
takes place, while the coupling to H$_g$ phonons decreases $U_c$.
Without electron-phonon coupling the Hund's rule coupling also
decreases $U_c$. There is, however, an interesting competition 
between the Hund's rule coupling and the Jahn-Teller effect.
Thus the reduction of $U_c$ due to the H$_g$ phonons becomes
smaller when the Hund's rule coupling is turned on. The implications 
of this is discussed in the context of A$_n$C$_{60}$.
\end{abstract}
\begin{multicols}{2}
\section{Introduction}
The alkali-doped fullerenes, A$_n$C$_{60}$ (A= K, Rb), raise many 
interesting problems due to the important role played by both the 
electron-electron and the electron-phonon interaction\cite{RMP}. 
For $n=3$ and $n=4$ each alkali atom is assumed to donate 
about one electron into a six-fold degenerate $t_{1u}$ band. 
A$_3$C$_{60}$ are metals and superconductors\cite{Haddon,Hebard}, while 
A$_4$C$_{60}$ are insulators\cite{Murphy,Kiefl}. The $t_{1u}$ orbital
is only partly filled for both systems, and band structure calculations
predict both systems to be metallic\cite{ErwinA4}.
A$_4$C$_{60}$ must then be an insulator due to interaction
effects neglected in band structure calculations. 
Under pressure A$_4$C$_{60}$ becomes metallic\cite{Kerkoud},
while some fullerenes at normal pressure and with the doping $n=3$ 
(NH$_3$K$_3$C$_{60}$, Cs$_3$C$_{60}$) are not superconductors 
and probably insulators, but become superconductors under
pressure\cite{Iwasa,Zhou,Palstra}.
This suggest that these systems are relatively close to a metal-insulator
transition. A metal-insulator transition is usually discussed
in terms of the ratio between the Coulomb interaction $U$ between two
electrons on the same molecule and the one-particle band width $W$.
For A$_3$C$_{60}$ and A$_4$C$_{60}$ this ratio is, however, almost 
identical\cite{ErwinA4,C60A4}. 
It is then interesting to ask which factors determine
whether a system is on the metallic or insulating side.

These systems have been studied in a Hubbard-like model, and it was 
found that the three-fold degeneracy of the $t_{1u}$ orbital plays 
an important role by increasing the ratio $U/W$ where the metal-insulator 
transition takes place\cite{deg,Jong}.  It was furthermore found that 
the lattice structure is important for the difference between 
A$_3$C$_{60}$ (fcc) and A$_4$C$_{60}$ (bct)\cite{frust,frust1}. 
Hubbard-like models predict, however, that A$_4$C$_{60}$ is an 
anti-ferromagnetic insulator, while it is known experimentally 
that there are no moments in A$_4$C$_{60}$\cite{Kiefl,paramagnetic}. 
The electrons in A$_n$C$_{60}$ have a relatively strong interaction 
with Jahn-Teller intramolecular phonons with H$_g$ symmetry and a weaker 
interaction with A$_g$ phonons. The interaction with the Jahn-Teller 
phonons favor a low spin state and might lead to a nonmagnetic 
insulator\cite{Fabrizio}. At the same time there is, however, a
Hund's rule coupling, which favors  a high spin state. This leads to an
interesting competition between the Jahn-Teller effect and the 
Hund's rule coupling. The purpose of this paper is therefore to study
the influence of the Jahn-Teller effect and the Hund's rule coupling
on the metal-insulator transition. We do this in the context of the 
Fullerenes, but similar effects also occur in many other systems, e.g.,
transition metal compounds\cite{Imada}. 

In this paper we study a model of A$_n$C$_{60}$ which includes the 
electron-phonon coupling and the electron-electron interaction. In 
A$_n$C$_{60}$ partly occupied orbital   $t_{1u}$ 
couples to two nondegenerate phonons of A$_g$ symmetry and to 
eight five-fold degenerate (Jahn-Teller) phonons of H$_g$ symmetry. 
For A$_n$C$_{60}$ the important coupling is believed to be to the H$_g$
phonons\cite{RMP}, and we therefore study a model with a coupling to one
H$_g$ phonon. To see the effect of having a Jahn-Teller phonon, we also
compare with a model containing an A$_g$ phonon. The  models are 
solved in the dynamical mean-field theory (DMFT)\cite{DMFT} or by
using exact diagonalization. To interprete the results we also study 
analytically a single molecule and a simple two-site model. 
A brief summary of some aspects of this work are published
elsewhere\cite{frust}.

In Sec. \ref{sec2} we present results for a free molecule and in
Sec. \ref{sec2a} we discuss the typical parameter range.
We present results for a two-site model in Sec. \ref{sec3} and describe 
the DMFT calculation in Sec. \ref{sec4}.  The results are given in Sec. 
\ref{sec5} and discussed in Sec. \ref{sec6}.

\section{Isolated molecule}\label{sec2}
We first study an isolated molecule. We consider the case of
the coupling to an A$_g$ phonon and two cases of coupling to Jahn-Teller
phonons, namely the $E\times E$ case, where a two-fold 
degenerate level couples to a two-fold degenerate phonon, and
the $T\times H$ case, where a three-fold degenerate level
couples to a five-fold degenerate phonon. The $E\times E$ case
is the simplest model with a Jahn-Teller effect, and the $T\times H$ 
case is a simple model of a C$_{60}$ molecule. 
Thus we consider the Hamiltonian

\begin{eqnarray}\label{eq:1}
H_{\rm el-ph}&&=\varepsilon_0 \sum_{m} \sum_{\sigma} 
c^{\dagger}_{m\sigma}c_{m\sigma}
+\omega_{0}\sum_{\nu}b^{\dagger}_{\nu}b_{\nu} \nonumber \\ 
&&+ {g} \sum_{\nu} \sum_{\sigma}\sum_{m}
\sum_{m^{'}}
V_{mm^{'}}^{(\nu)}c^{\dagger}_{m\sigma} c_{m^{'}\sigma}(b_{\nu}+ 
b_{\nu}^{\dagger}),
\end{eqnarray}
where the first term describes the electronic level, the second the 
phonon and the third the electron-phonon coupling. 
The coupling matrices $V^{(\nu)}$ are determined by symmetry and they 
are given in, e.g., Ref. \onlinecite{jtbook,Lannoo,c60jt}.
For the coupling to an A$_g$ phonon $V^{(1)}$ is diagonal with all
the diagonal elements equal to unity, $V^{(1)}_{mm^{'}}=\delta_{mm^{'}}$. 
For the $E\times E$ case  
\begin{equation}\label{eq:2}
V^{(1)}=\left(\begin{array}{cc} 1 & 0 \\ 0& -1\end{array}\right)
\hskip0.5cm {\rm and} \hskip0.5cm
V^{(2)}=\left(\begin{array}{cc} 0 & 1 \\ 1& 0\end{array}\right)
\end{equation}
and for the $T\times H$ case
\begin{eqnarray} \label{eq:3}
&&V^{(1)}={1\over 2}\left(\begin{array}{ccc}
-1 \ & 0  & 0 \\
0  & -1 \ & 0 \\
0  &  0 & 2 \ \end{array}\right)\  \hskip0.0cm 
V^{(2)}={1\over 2}\left(\begin{array}{ccc}
\sqrt{3} & 0  & 0 \\
0  & -\sqrt{3} & 0 \\
0  &  0 & 0 \end{array}\right)\   \nonumber \\
&&V^{(3)} \ =\ {\sqrt{3}\over 2}\left(\begin{array}{ccc}
0  & 1  & 0 \\
1  & 0  & 0 \\
0  &  0 & 0 \end{array}\right)\ \hskip0.4cm 
V^{(4)} \ = \ {\sqrt{3}\over 2}\left(\begin{array}{ccc}
0  & 0  & 1 \\
0  & 0  & 0 \\
1  &  0 & 0 \end{array}\right)\   \\
&&V^{(5)} \ = \ {\sqrt{3}\over 2}\left(\begin{array}{ccc}
0  & 0  & 0 \\
0  & 0  & 1 \\
0  &  1 & 0 \end{array}\right)\ \nonumber
\end{eqnarray}

The overall coupling strength is determined by $g$. In a metallic molecular 
solid with narrow bands, there is a simple relation between $g$ and
the dimensionless coupling $\lambda$\cite{Lannoo}
\begin{equation}\label{eq:5}
\lambda_{H}={5\over 3}N(0){g^2\over \omega_0}
\end{equation}
for the $T\times H$ case and 
\begin{equation}\label{eq:4}
\lambda_{E}=2N(0){g^2\over \omega_0}
\end{equation}
for the $E\times E$ case.                  
Here  $N(0)$ is the electron density of states per spin.      
We also define a Jahn-Teller energy
\begin{equation}\label{eq:6}
E_{JT}={g^2\over \omega_0}.
\end{equation} 

In addition we include the Coulomb interaction
\begin{eqnarray}\label{eq:7}
H_{\rm U}=&&U_{xx}\sum_{m} n_{m\uparrow}n_{m\downarrow}
+U_{xy}\sum_{\sigma\sigma^{'}}\sum_{m<   m^{'}}n_{\sigma m}
n_{\sigma^{'}m^{'}}  \nonumber  \\
+&&{1\over 2}K \sum_{\sigma\sigma^{'}}\sum_{m\ne m^{'}}
\psi^{\dagger}_{m\sigma}\psi^{\dagger}_{m^{'}\sigma^{'}}
\psi_{m\sigma^{'}}\psi_{m^{'}\sigma }   \\
+&&{1\over 2}K \sum_{\sigma}\sum_{m\ne m^{'}}
\psi^{\dagger}_{m\sigma }\psi^{\dagger}_{m-\sigma }
\psi_{m^{'}-\sigma }\psi_{m^{'}\sigma },  \nonumber  
\end{eqnarray}
where $n_{m\sigma}=\psi^{\dagger}_{m\sigma}\psi_{m\sigma}$
is an occupation number, $U_{xx}$ and $U_{xy}$ are the Coulomb 
interactions between equal and unequal orbitals, respectively 
and $K$ is an exchange integral. The Coulomb integrals are 
related via
\begin{equation}\label{eq:8}
U_{xy}=U_{xx}-2K.
\end{equation}

For A$_g$ phonons and a free molecule, the electron-phonon problem is 
reduced to a displaced oscillator, and we can easily obtain the 
electron-phonon contribution to the ground-state energy
\begin{equation}\label{eq: 10}
E^{A_g}(N)=-N^2E_{JT}.              
\end{equation}
where $E(N)$ is the energy of a free molecule with $N$ electrons.
The Jahn-Teller case is more complicated and cannot be solved 
exactly. We therefore focus on the lowest order contribution to 
$E(N)$. For the Coulomb interaction 
this contribution is of first order in $K$ and for the electron-phonon 
interaction it is  of second order in $g$. The results are shown in Table 
\ref{table1} for the $E\times E$ case and in Table \ref{table2} for 
the $T\times H$ case. We have subtracted an average Coulomb energy,         
\begin{equation}\label{eq:9}
\tilde E(N)\equiv E(N)-{1\over 2}N(N-1)U_{av}(k),
\end{equation}
where $U_{av}(k)$ is the average interaction for the ground-state in a 
full shell with $2k$ electrons and the orbital degeneracy $k$. 
By definition, $\tilde E(N)$ is then zero for a full shell. 
Results for the low spin state (singlet or doublet) and the high spin
state (triplet or quartet) are shown in Table \ref{table1}.

Neglecting the Coulomb interaction ($U$ and $K$) the results 
for the $T\times H$ case can be written as\cite{Auerbach,c60jt}
\begin{equation}\label{eq:9a}
E_H^{WC}(N)={5\over 2}C(N)E_{JT},
\end{equation}
where 
\begin{equation}\label{eq:9b}
C(N)=\left\{\begin{array}{ccc}
1 & {\rm for} & N=1, 5  \\
4 & {\rm for} & N=2, 4  \\
3 & {\rm for} & N=3                 
\end{array}\right.
\end{equation}
This is a weak-coupling result for the case when $\lambda$ (or $g$)
is small. For the strong-coupling case\cite{Auerbach,c60jt}
\begin{equation}\label{eq:9c}
E_H^{SC}(N)=C(N)E_{JT}.
\end{equation}
In the strong-coupling case the prefactor in front of $C(N)E_{JT}$
has been reduced by a factor of $5/2$.

From tables \ref{table1} and \ref{table2} it follows that
the low spin state is lower in energy for $K<2E_{JT}$ in the $E\times E$ 
case and for $K<{3\over 2}E_{JT}$ in the $T\times H$ case.
The results illustrate the competition between the 
Jahn-Teller effect and the Hund's rule coupling. According to 
Hund's first rule, it is favorable to form a high spin state to
minimize the Coulomb energy. On the other hand, the Jahn-Teller effect
favors a low spin state. For instance, if the first phonon ($\nu=1$) 
of $E$ symmetry (the left matrix in Eq. (\ref{eq:2})) is excited, it 
is favorable to put two electrons (with opposite spins) in the level 
$m=2$, resulting in a singlet. The tables show that in the low spin 
state the energy $\tilde E(N)$ is lowered by the electron-phonon 
coupling but is increased by the exchange integral $K$. 
In the high spin state, on the other hand, the exchange coupling lowers 
the energy, while the electron-phonon interaction lowers the energy 
less than in the low spin state or not at all.  

This can be further illustrated by considering the case of two 
electrons in the $E\times E$ system. For $K<2E_{JT}$, the important 
states are 
\begin{equation}\label{eq:10}
|1\rangle=\psi^{\dagger}_{1\uparrow}\psi^{\dagger}_{1\downarrow}|
{\rm vac}\rangle \hskip1cm 
|2\rangle=\psi^{\dagger}_{2\uparrow}\psi^{\dagger}_{2\downarrow}|
{\rm vac}\rangle.         
\end{equation}
These states couple directly via the exchange interaction and 
indirectly via the electron-phonon interaction. Thus we also
include states where one phonon has been excited. In the corresponding
Hamiltonian matrix the part corresponding to the states with one phonon
are ``folded'' into the part corresponding to $|i\rangle$, $i=1$ or 2,
and we obtain the matrix
\begin{equation}\label{eq:11}
\left(\begin{array}{cc} U_{xx}-6E_{JT}  &  -2E_{JT}+K  \\
 -2E_{JT}+K & U_{xx}-6E_{JT}  \end{array} \right)
\end{equation}
If $K<2E_{JT}$ the corresponding energy is           
\begin{equation}\label{eq:12}
U_{xx}-6E_{JT}+(-2E_{JT}+K)=U_{av}(2)-8E_{JT}+{8\over 3}K,
\end{equation}
which clearly shows the competition between the Jahn-Teller 
coupling and the Hund's rule coupling. For $K>2E_{JT}$ a triplet
state of the type $\psi^{\dagger}_{1\uparrow}\psi^{\dagger}_{2\uparrow}|
{\rm vac}\rangle$ with the energy $U_{av}(2)-4K/3$ becomes the lowest
state.

\section{Parameter range}\label{sec2a}
\begin{figure}[bt]
\centerline{
\rotatebox{270}{\resizebox{!}{3in}{\includegraphics{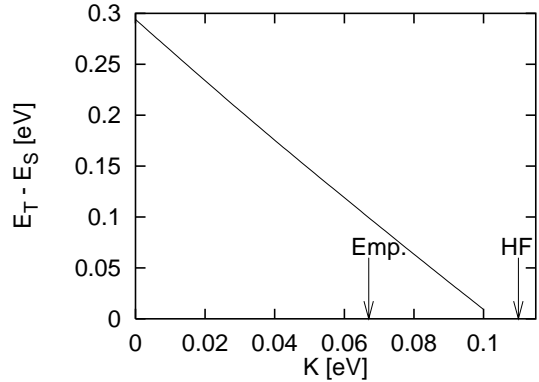}}}}
\caption[]{\label{fig:splitting}The triplet-singlet splitting of a 
C$_{60}^{-4}$ molecule with one H$_g$ phonon and an exchange 
integral $K$. The arrows show the value of $K$ deduced from a 
HF calculation and the empirical value deduced from the experimental
triplet-singlet splitting.}
\end{figure}

It is now important to establish the parameter range appropriate 
for A$_n$C$_{60}$. In particular, the relative size of $g$ and
$K$ is important, since this determines whether the Jahn-Teller 
effect or the Hund's rule coupling wins. The exchange integral
$K$ has been estimated\cite{K} from an  {\it ab initio} SCF 
calculation\cite{correlated}. From the calculated multiplet splitting        
the result $K=0.11$ eV was obtained. Correlation effects are expected
to reduce this number, since correlation effects in particular lower the 
energies of the low spin states. The difference in Coulomb energy
between the low spin states and the high spin states is then reduced
and the effective $K$ becomes correspondingly smaller. For instance, 
for some atomic multiplets, correlation effects were found to
reduce the multiplet splitting by about 25 $\%$\cite{Barth}.
The electron-phonon coupling constants have been 
estimated from photoemission experiments for a free molecule\cite{PES}. 
Here we replace the eight H$_g$ modes by one effective
mode. The frequency of this mode is chosen as the logarithmically 
averaged frequency
\begin{eqnarray}\label{eq:12a}
\lambda =\sum_{\nu=1}^8 \lambda_{\nu}; \hskip0.5cm
\lambda \ {\rm ln} \omega_{0}=\sum_{\nu=1}^8 \lambda_{\nu}\ {\rm ln} 
\omega_{\nu}
\end{eqnarray} 
where $\lambda_{\nu}$ and $\omega_{\nu}$ are the electron-phonon
couplings and frequencies, respectively.
We have calculated the energies of the lowest singlet and triplet
states for a free C$_{60}$ molecule. The results are shown in
Fig. \ref{fig:splitting}. Experimentally it is found that  
the low spin state wins for both  A$_3$C$_{60}$ and A$_4$C$_{60}$,
A$_4$C$_{60}$ being a nonmagnetic insulator\cite{Kiefl,paramagnetic}
and NH$_3$K$_3$C$_{60}$ being antiferromagnetic with a moment 
(0.7 $\mu_B$ per molecule)\cite{Prassides} which 
corresponds to a spin 1/2 system.
For A$_4$C$_{60}$ the triplet-singlet splitting is estimated to be    
0.1 eV\cite{paramagnetic}. In Fig. \ref{fig:splitting} this splitting is
obtained for $K\sim 0.07$ eV. As expected this value is smaller 
than the value $K=0.11$ eV deduced from the HF calculation.
We observe that a change in the estimate of $g$ of course would also lead 
to a change in this empirical number of $K$. Since, however, the
present estimate of $K$ seems reasonable relative the the HF value, 
this calculation also gives some support for the value of $g$ 
deduced from PES.

\section{A two-site model}\label{sec3}
In this section we study a two-site model to gain understanding of 
a) the different influences of an A$_g$ and a Jahn-Teller 
phonon on the metal-insulator transition and b) the competition between
the Jahn-Teller effect and the Hund's rule coupling. We therefore compare the   
case of an A$_g$ phonon coupling to a two-fold degenerate level
($A\times E$ problem) with the Jahn-Teller $E\times E$ problem. 
Thus we study the model
\begin{equation}\label{eq:a0}
H=\sum_{i=1}^2\lbrack H_{\rm el-phon}(i)+H_{\rm U}(i) \rbrack +H_{\rm hop},
\end{equation}
where $H_{\rm el-phon}(i)$ (Eq. (\ref{eq:1})) describes the electron-phonon 
interaction on site $i$, $H_{\rm U}(i)$ (Eq. (\ref{eq:7})) describes the 
Coulomb interaction on site $i$, and 
\begin{equation}\label{eq:a1}
H_{\rm hop}=-t\sum_{m=1}^2\sum_{\sigma}(\psi^{\dagger}_{1m\sigma}\psi_{2m\sigma}
+\psi^{\dagger}_{2m\sigma}\psi_{1m\sigma}),               
\end{equation}
describes the hopping between the two sites. 
We have assumed that there is only hopping between equal $m$-quantum numbers.

The band gap $E_g(n)$ for the filling $n$ is given by
\begin{equation}\label{eq:a3}
E_g(n)=E(2n-1)+E(2n+1)-2E(2n)
\end{equation}
for a two-site model, where $E(N)$ is the ground-state energy
of a system with $N$ electrons.
It is useful to consider $E_g$ in the limit when $U$ is very large, and then
to extrapolate to intermediate values of $U$. The limit of a large $U$ is 
particular transparent, allowing for simple calculations and a qualitative
understanding, but it is still relevant for the intermediate values of $U$
where the metal-insulator transition takes place\cite{deg}.
The the metal-insulator transition happens when   
$E_g=0$, i.e., we find the critical value of $U$ for which this
condition is satisfied. The two-site system is
much too small for obtaining reliable quantitative estimates of
the critical $U$ and the extrapolation of the large $U$ results
is questionable. Nevertheless, this approach can give a qualitative 
understanding of more realistic calculations\cite{deg}.  We consider the limit 
\begin{eqnarray}\label{eq:a4}
&&g << \omega_0 << W<< U \nonumber \\
&&K\sim {g^2\over \omega_0} \equiv E_{JT}<< W,
\end{eqnarray}
where the electron-phonon coupling and the exchange integral are just 
weak perturbations to the hopping and to the Coulomb interaction.  

We first study the case of an A$_g$ phonon and calculate the
gap for the filling $n=1$. It is then convenient to transform the
coupling term to the form
\begin{equation}\label{eq:a5}
g\sum_{i=1}^2 (n_i-1)(b_i+b_i^{\dagger}),
\end{equation}
where $n_i=\sum_{m\sigma}\psi^{\dagger}_{im\sigma}\psi_{im\sigma}$
and we have neglected an irrelevant term $-(g^2/\omega_0)\sum_i(2n_i-1)$.
For a system with two electrons and in the limit of a large $U$, hopping
is almost completely suppressed, and each site has 
almost exactly one electron. The term (\ref{eq:a5}) then has 
essentially zero coupling to this state, and we obtain
\begin{equation}\label{eq:a6}
E^{A_g}(2)\approx 0.
\end{equation}
Since there is only one electron per site, there are no 
multiplet effects.
To obtain the energy of a system with one electron, it is convenient 
to transform to bonding and antibonding operators, e.g., 
$\psi_{\pm m\sigma}=(\psi_{1m\sigma}\pm \psi_{2m\sigma})$. This gives 
the interaction term
\begin{eqnarray}\label{eq:a7}
{g\over \sqrt{2}}\lbrack &&(\psi^{\dagger}_{+m\sigma}\psi_{+m\sigma}+
\psi^{\dagger}_{-m\sigma}\psi_{-m\sigma})(b_++b^{\dagger}_+)\nonumber \\
+&& (\psi^{\dagger}_{+m\sigma}\psi_{-m\sigma}+
\psi^{\dagger}_{-m\sigma}\psi_{+m\sigma})(b_-+b^{\dagger}_-)\rbrack.
\end{eqnarray}
The bonding level is then occupied by one electron, giving the 
kinetic energy $-t$. The leading contribution to the energy due to
the electron-phonon interaction is $-{1\over 2}E_{JT}$, giving
\begin{equation}\label{eq:a8}
E^{A_g}(1)=-t-{1\over 2}E_{JT}.               
\end{equation}
In the limit studied here ($\omega_0<<W$), the phonons react to the 
average electronic charge, which gives rise to the factor ${1\over 2}$.

We next consider the system with three electrons.        
The wave function for $g=0$ and $K=0$ is
\begin{eqnarray}\label{eq:a9}
&&|0\rangle={1\over \sqrt{6}}(\psi^{\dagger}_{11\uparrow}
\psi^{\dagger}_{11\downarrow}\psi^{\dagger}_{22\uparrow}+
\psi^{\dagger}_{11\uparrow}
\psi^{\dagger}_{21\downarrow}\psi^{\dagger}_{12\uparrow}+
\psi^{\dagger}_{21\uparrow}
\psi^{\dagger}_{11\downarrow}\psi^{\dagger}_{12\uparrow}\nonumber \\
+&& \psi^{\dagger}_{21\uparrow}
\psi^{\dagger}_{21\downarrow}\psi^{\dagger}_{12\uparrow}+
\psi^{\dagger}_{21\uparrow}
\psi^{\dagger}_{11\downarrow}\psi^{\dagger}_{22\uparrow}+
\psi^{\dagger}_{11\uparrow}
\psi^{\dagger}_{21\downarrow}\psi^{\dagger}_{22\uparrow}
)|{\rm vac}\rangle.
\end{eqnarray}
The corresponding energy is $\langle 0|H|0\rangle=-2t$. For a general
lattice and for a state which is half-filled apart from one electron 
or one hole, we expect the hopping to be enhanced by roughly a factor 
$\sqrt{N_{\rm deg}}$, where $N_{\rm deg}$ is the orbital 
degeneracy\cite{deg}. However, for the two-site system this enhancement
is $N_{\rm deg}$\cite{deg}. The effects of the electron-phonon coupling 
are now included in perturbation theory. For the parameter
range considered here ($\omega_0<<W$), there is only coupling
to the states $b^{\dagger}_i|0\rangle$. Furthermore, the Coulomb energy 
is given by $\langle 0|H_{\rm U}|0\rangle$, since there are no states with 
the kinetic energy -$2t$ which couple to $|0\rangle$ via $H_{\rm U}$.
The energy is therefore
\begin{equation}\label{eq:a10}
E^{A_g}(3)=U_{xx}-2t-{1\over 2}E_{JT}-{5 \over 3}K.
\end{equation}
The corresponding band gap is given in Table \ref{table3} ($A\times E$),
which also shows the result for the filling $n=2$.
The table shows that the band gap is reduced by the electron-phonon 
coupling. Extrapolating to intermediate values of $U$ then suggests 
that the electron-phonon coupling increases the critical 
value of $U$ where the metal-insulator transition takes place. 
The reason is that the electron-phonon interaction in the model 
(\ref{eq:a5}) is not effective for integer filling in the large $U$ case, 
since the charge
fluctuations are then suppressed. For the states with an extra            
electron or hole, on the other hand, there is a fluctuating charge
due to the hopping of the electron or hole and a corresponding lowering 
of the energy due to the electron-phonon coupling. 

The multiplet effects trivially reduce the gap for $n=1$, since they 
are effective for the state with three electrons but 
not for the other two states. For $n=2$, however, they increase the gap. 
The reason is that in the integer occupation case ($N=4$), hopping plays no  
role for a large $U$ and the state can adjust to use the multiplet
effects optimally. This is not possible for the case of an extra hole
($N=3$) or an extra electron ($N=5$), since then the states first of
all adjusts to optimize hopping, and only in the second place adjust
to optimize the multiplet effects.  The increase of the gap due to the 
exchange integral has been observed earlier\cite{Jong}.

We next consider the Jahn-Teller case of two-fold degenerate phonons.
We first calculate the energy of the state with one electron per
site in the large $U$ limit. Perturbation theory shows that each
phonon contributes an energy $E_{JT}$. Since there are
two phonons per site and two sites, the energy is
\begin{equation}\label{eq:a13}
E^{E_g}(2)=-4E_{JT},
\end{equation}
where E$_g$ labels the Jahn-Teller case. Although charge fluctuations 
are completely suppressed in this case, the system can still gain
energy by introducing a (dynamical) Jahn-Teller distortion. This
is in strong contrast to the A$_g$ phonons, which only couple
to the net charge on a given site. In perturbation theory,
the energies of the states with an extra electron or hole are
\begin{eqnarray}\label{eq:a14}
&&E^{E_g}(1)=-t-E_{JT}  \nonumber \\ 
&&E^{E_g}(3)=U_{xx}-2t-E_{JT}-{5\over 3}K. 
\end{eqnarray} 
The electron-phonon interaction lowers the energy twice as much as 
in the A$_g$ case (Eqs. (\ref{eq:a8}) and (\ref{eq:a10})), simply because 
the phonon is two-fold degenerate and the coupling therefore is to twice 
as many phonons as in the A$_g$ case.
The corresponding gap is shown in Table \ref{table3}, which also shows the 
gap for the filling $n=2$.

For $n=2$ and $K<E_{JT}$ the Jahn-Teller effect dominates over the Hund's rule
coupling. In this case as well as for $n=1$ the gap is then increased 
by the electron-phonon interaction, while in the A$_g$ case it is 
decreased. The reason is similar as for the multiplet effects in
the A$_g$ case, discussed in the second paragraph below Eq. (\ref{eq:a10}).
For the parameter range (Eq. (\ref{eq:a4})) we are considering and 
for integer filling, the hopping is very efficiently suppressed, and the 
Jahn-Teller system can therefore adjust efficiently to the 
electron-phonon interaction. With an extra electron or hole, however, 
the system cannot take advantage of the electron-phonon interaction 
to the same extent, since the wave function primarily optimizes the
hopping of the electron or hole. According to Eq. (\ref{eq:a3}),
 this leads to an increase of the gap. 
In the A$_g$ case, on the other hand, 
the system cannot couple to the phonons in the integer filling case,
and therefore even the reduced coupling to the phonons in the case 
of an extra electron or hole is sufficient to reduce the gap.

Table \ref{table3} furthermore illustrates the competition between 
the Jahn-Teller effect and the Hund's rule coupling.  For $K<2E_{JT}$  
and $n=2$ an increase of $K$ leads to a reduction of the gap.
As $K$ is further increased, the system instead tends to go into
high spin states, and the gap is increased as K is increased. 

It is interesting to compare the results with a different approach.
We can calculate an effective on-site $U_{eff}(n)$ for the filling 
$n$  as
\begin{equation}\label{eq:a17}
U_{eff}(n)=E(n+1)+E(n-1)-2E(n),
\end{equation}
where $E(N)$ now refers to the energy of a free molecule, calculated 
in Sec. \ref{sec2}. For $K<2E_{JT}$ and for the $E\times E$ case,
 we then obtain 
\begin{eqnarray}\label{eq:a18}
&&U_{eff}(1)=U_{av}(2)-4 E_{JT}+{8\over 3}K \nonumber \\
&&U_{eff}(2)=U_{av}+12E_{JT}-{16\over 3}K.
\end{eqnarray}
It is then tempting to assume that for the filling $n=1$ and $n=2$ 
the gap is given by $U_{eff}(1)-3t$ and $U_{eff}(2)-4t$, respectively. 
Comparison with Table \ref{table3} shows that this 
is incorrect in the limit studied here (Eq. (\ref{eq:a4})). 
In particular, for $n=1$, this approach would even predict 
the wrong sign for the the electron-phonon contribution to the gap. 
The reason is that Eq. (\ref{eq:a17}) assumes
that the electron-phonon and Hund's rule couplings can adjust 
to the instantaneous occupation of a given site. As we have seen 
above, this is not possible in the limit (Eq. (\ref{eq:a4}))
for the states with an extra electron or hole.  
$U_{eff}(2)$ may instead become relevant
in the limit when the exchange coupling $K$ and the Jahn-Teller 
energy $E_{JT}$ are much larger than the hopping energy. 

In a similar way we have calculated the gap for the $T\times H$
two-site problem. The results are shown in Table \ref{table4}. 
These results also illustrate the competition between the
Jahn-Teller effect and the Hund's rule coupling.

\section{Dynamical mean-field calculations}\label{sec4}

We now consider the $T\times H$ problem in the dynamical mean-field
theory (DMFT).
We formulate the infinite dimensional limit on the Bethe lattice,
where the nearest-neighbor hopping integral $t_{im,jm'}$ is rescaled 
as $t^*/\sqrt{z}$ with the connectivity $z$ going to infinity. We
further simplify the hopping by setting $t_{im,jm'}\propto\delta_{mm'}$ 
only allowing the diagonal hopping. The unit of energy is chosen such
that the bandwidth, $W$, is set to 2. 
The Jahn-Teller phonons embedded at each lattice sites of the bath
can be easily incorporated into the effective medium of
the impurity Anderson model\cite{DMFT_Hol}, 
since they are Einstein phonons without direct intersite
couplings between them. Therefore the lattice contribution of phonons
is implicitly included in the medium electron Green's function. 

We here focus on the problem where the exchange integral $K=0$.
The effective impurity Anderson model is solved using the quantum Monte
Carlo (QMC) technique with the Fye-Hirsch algorithm\cite{H-F}, which has
only mild ``fermion sign-problems'' with our Hamiltonian. Here we
treat fully quantum mechanically the phonon fields which are updated
together with the fermion auxiliary fields in each Monte Carlo step. 
Details of the implementation of the QMC technique in the DMFT can be
found elsewhere in the literature\cite{DMFT,DMFT_Hol}.
We have used the discretization step for the Trotter breakup, 
$\Delta\tau=1/3$, throughout this paper unless mentioned otherwise and
more than one million Monte Carlo sweeps are taken for each iteration
of the self-consistency loop. 
For the case where the exchange integral $K>0$ the QMC method has a 
serious ``sign-problem'', and this case is treated in Sec. \ref{sec5}D
using exact diagonalization.

\section{Results}\label{sec5}
We study how the metal-insulator transition depends on the
parameters, e.g., the ratio $U/W$ for different strengths 
of the electron-phonon coupling $\lambda$ and the exchange integral.
When the system becomes insulating a gap is opened up in the 
electron spectral function $A(\omega)$. This shows up in the electron
Green's function $G(\tau)$ calculated for imaginary times $\tau$. 
For instance if 
\begin{equation}\label{eq:d1}
A(\omega)=0 \hskip 1cm {\rm for} \hskip1cm |\omega|<\Delta,
\end{equation}
we obtain
\begin{equation}\label{eq:d2}
G(\tau=\beta/2)\le e^{-\Delta \beta/2},
\end{equation}
where $\beta=1/T$. Thus $G(\beta/2)$ decays exponentially with $\beta$
for an insulator. We therefore use the behavior of $G(\beta/2)$ as a 
measure whether the system is a metal or an insulator. 
It is also interesting to
study the charge fluctuation $\langle (n-n_0)^2\rangle$, 
which is an average of $\langle (n_i-n_0)^2\rangle$ and where
$n_0$ is the average occupancy per site.
This quantity is expected to become small but nonzero 
at the metal-insulator transition.

\subsection{Jahn-Teller H$_g$ phonons}
\begin{figure}[bt]
\centerline{
\rotatebox{270}{\resizebox{!}{3in}{\includegraphics{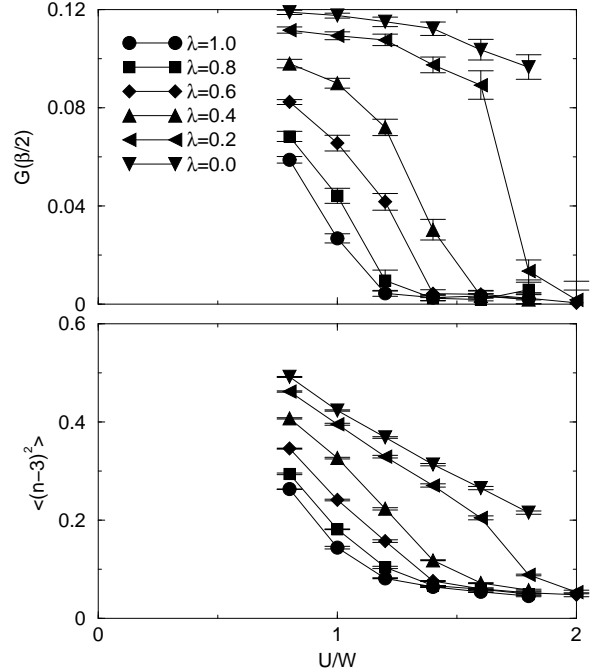}}}}
\caption[]{\label{fig1}(a) $G(\beta/2)$ for $\beta=16$ and (b) the charge 
fluctuation $\langle (n-3)^2\rangle$ as a function of $U/W$ 
for different values of $\lambda$ and coupling to H$_g$ phonons
in A$_3$C$_{60}$.}
\end{figure}

We first study the Jahn-Teller H$_g$ phonons and consider the case 
of half-filling, i.e., three electrons per site.
Fig. \ref{fig1} shows $G(\beta/2)$                                     
as a function of $U/W$ for different values of $\lambda$. 
The figure illustrates that for a given $\lambda$, $G(\beta/2)$ is 
reduced as $U$ is increased and at some critical value $U_c$,  
$G(\beta/2)$ becomes very close to zero (not exactly equal to 
zero due to the finite temperature), where a metal insulator 
transition takes place. The figure also illustrates that the 
charge fluctuations $\langle (n-3)^2\rangle$ are strongly reduced 
in the insulating state. The critical $U_c/W$ is reduced as $\lambda$ 
grows. The reason for this was discussed extensively in Sec. \ref{sec3}.

It is interesting that $U_c$ as a function of $\lambda$ 
initially is reduced very strongly as $\lambda$ is increased,
while for larger values of $\lambda$ the decrease is slower. 
This can be understood in terms of the results in Eq. (\ref{eq:9a},
\ref{eq:9c}) for the Jahn-Teller energy in the weak- and strong-coupling 
limits. This shows that the electron-phonon energy increases much
faster with $\lambda$ (by a factor of $5/2$) in the weak-coupling 
limit than in the strong-coupling limit. This is particularly
relevant for the large $U$ integer filling case, where the electron-phonon
interaction has a similar effect as in the free molecule, and it gives
a qualitative explanation for the dependence of $U_c$ on $\lambda$.

\begin{figure}[bt]
\centerline{
\rotatebox{270}{\resizebox{!}{3in}{\includegraphics{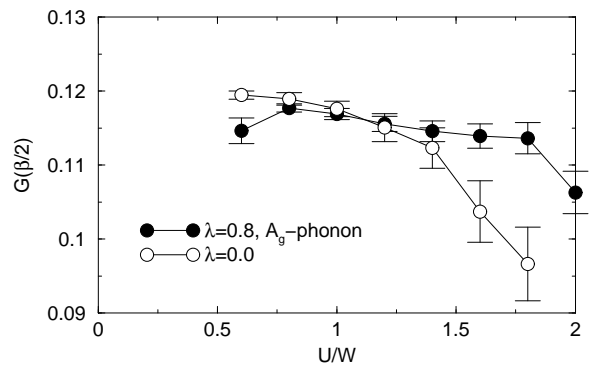}}}
}
\caption[]{\label{fig2} $G(\beta/2)$ for $\beta=16$ as a function 
of $U/W$ for $\lambda=0.8$ and $\lambda=0$ and coupling to A$_g$ 
phonons in A$_3$C$_{60}$.}                 
\end{figure}
\subsection{A$_g$ versus Jahn-Teller H$_g$ phonons}
We next compare the Jahn-Teller H$_g$ phonons with A$_g$ 
phonons. Fig. \ref{fig2} shows $G(\beta/2)$ in a system with 
coupling to A$_g$ phonons as a function of $U$ 
for $\lambda=0$ and for $\lambda=0.8$. Comparing the results for 
$\lambda=0$ and $\lambda=0.8$, we can see that the coupling 
to A$_g$ phonons increases the critical $U_c$ where the metal-insulator
transition takes place, while in the case of H$_g$ phonons this 
value is decreased (see Fig. \ref{fig1}). The reason for this 
change was discussed extensively 
in Sec. \ref{sec3}. 

It is interesting to observe that both the electron-phonon interaction 
and the electron-electron interaction by themselves would tend to
reduce $G(\beta/2)$. In a weak-coupling theory where we simply add
the lowest order contribution to the self-energy from each interaction,
we would then predict that the two interactions work together in 
reducing $G(\beta/2)$. This is indeed what is found in Fig. \ref{fig2}
for $U/W=0.5$. Not too surprisingly, the arguments presented below 
Eq. (\ref{eq:a14}) and assuming the large $U$ limit are incorrect
for $U/W=0.5$. It is not surprising that these arguments become
correct for large values of $U/W$, but it is important that they 
are qualitatively correct already for $U/W\sim 1.5$, which is on the 
metallic side of the metal-insulator transition.

For a system with only a Coulomb interaction $U$,
the metal-insulator transition can be thought of as resulting
from the competition between kinetic and potential energy.
When the coupling to the phonons is introduced the hopping 
of the electrons is reduced, since an  electron tends to 
drag a cloud of phonons along its path. It is then tempting 
to assume that the coupling to phonons will move the 
metal-insulator transition towards smaller values of $U$.  
Our results show that for the parameter range considered here,
the effect is the opposite for the case of coupling to 
A$_g$ phonons.

The coupling to the A$_g$ phonons in A$_3$C$_{60}$ is weak\cite{RMP},
and the A$_g$ phonons should not play an important role. There is,
however, a strong coupling to a charge carrying plasmon derived
from the $t_{1u}$ electrons\cite{plasmon}. This plasmon couples
in the same way as the A$_g$ phonons. Its energy (0.5eV) is larger        
than the phonon energies. However, whether we consider the 
plasmon energy to be small or large we arrive at the conclusion 
that it increases $U_c$. For small values of the plasmon energy this 
follows from the study of the A$_g$ phonons, and for large energies
of the plasmon, we can introduce an effective $U_{eff}$ as
in Eq. (\ref{eq:a17}). The coupling to the plasmon reduces 
$U_{eff}$ and therefore the metal-insulator transition happens
for a larger bare $U$. 

\subsection{A$_3$C$_{60}$ versus A$_4$C$_{60}$}
\begin{figure}[bt]
\centerline{
\rotatebox{270}{\resizebox{!}{3in}{\includegraphics{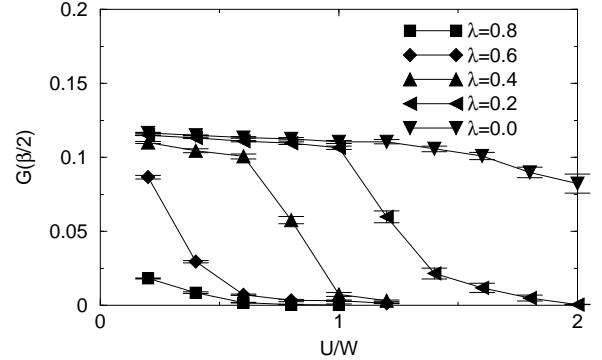}}}
}
\caption[]{\label{fig3} $G(\beta/2)$ for $\beta=16$ as a function 
of $U/W$ for A$_4$C$_{60}$ and different values of $\lambda$.} 
\end{figure}

Fig. \ref{fig3} shows $G(\beta/2)$ for A$_4$C$_{60}$. The figure
illustrates that $U_c$ is smaller at filling four than filling three 
(see Fig. \ref{fig3}). This is not surprising in view of Eq. 
(\ref{eq:9a}-\ref{eq:9c}) and Table \ref{table4}, since the energy
is lowered more by the electron-phonon coupling for filling
four than for filling three. According to Eq. (\ref{eq:a3}) this
increases the gap of the insulating state more for filling four than 
for filling three, i.e., it makes $U_c$ smaller for filling four.
As discussed in the next section, this result is, however, modified
by the Jahn-Teller coupling.

\subsection{Competition between Jahn-Teller and Hund's rule coupling}

\begin{figure}[bt]
\centerline{
\rotatebox{270}{\resizebox{!}{3in}{\includegraphics{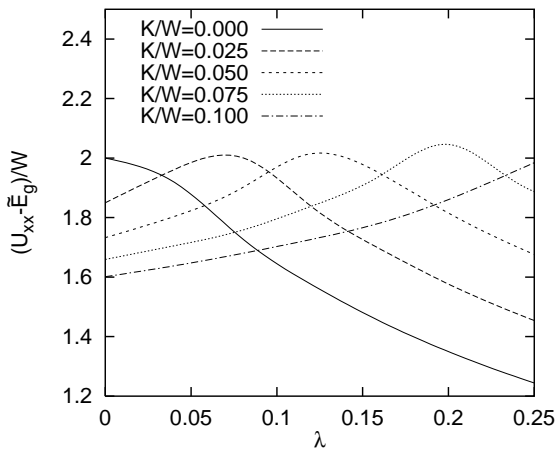}}}}
\caption[]{\label{fig:k}The simple estimate $U_{xx}-E_g^{red}$ of the    
critical $U_{xx}$ as a function of $\lambda$ for different values of the 
exchange integral $K$. The one-particle band width is $W=2$ and 
the phonon energy is $\omega_0=0.5$. For $\lambda$ we use 
$\lambda=4g^2/(\omega_0W)$.}
\end{figure}

We finally discuss the competition between the Jahn-Teller effect
and the Hund's rule coupling. Since there is a sign-problem 
in the dynamical mean-filed calculations if we use         
the full multiplet coupling in Eq. (\ref{eq:7}), we use an exact
diagonalization technique. This requires that the system size is small.
Therefore we consider a system with just four sites and we consider
the $E\times E$ problem (Eq. (\ref{eq:2})). We furthermore limit the 
Hilbert space by not allowing more than two phonons per site.
To reduce the discreteness of the one-particle spectrum for such 
a small system, we pick the hopping integrals randomly.
We calculate the gap according to Eq. (\ref{eq:a3}) and add some finite
size corrections\cite{deg}
\begin{equation}\label{eq:d3}
E_g^{red}(U_{xx})=E_g-{U_{av}(2)\over N_{site}}-E_g^{U=0},
\end{equation}
where $U_{av}(2)/N_{site}$ is a contribution to the gap from the electrostatic
energy of a finite system, $N_{site}$ is the number of sites and
$E_g^{U=0}$ is the gap of a system 
without any Coulomb interaction. Both these corrections go to zero
for a large system. Assuming that that $E_g^{red}$ grows linearly
with $U_{xx}$, we can then use 
\begin{equation}\label{eq:d4}
U_{xx}-E_g^{red}(U_{xx})
\end{equation}
as a crude estimate of the critical $U_{xx}$.
This quantity is shown in Fig. \ref{fig:k}.
The results can be qualitatively understood from the results in Table 
\ref{table3} for a two-site system, although the parameter range 
considered here is outside the range where the results in the table 
are valid.
For $\lambda=0$, the critical value of $U_{xx}$ is reduced as $K$
is increased, as expected\cite{Jong}. However, as $\lambda$ is 
increased, the competition between the Jahn-Teller and Hund's rule effects
leads to an increase of the critical $U_{xx}$. This is in agreement with
Table \ref{table3}, although the increase is faster than for the parameter
range of this table. For a particular value of $\lambda$,
the critical $U_{xx}$ becomes comparable to the value for $K=\lambda=0$. 
As $\lambda$ is further increased, the Jahn-Teller effect becomes 
dominating and the critical value of $U_{xx}$ decreases again. This 
critical value is, however, larger than for $K=0$, due to the 
competition between the Jahn-Teller and Hund's rule coupling.
\section{Discussion}\label{sec6}
It is interesting to discuss these results in the context of 
A$_n$C$_{60}$. Considering just the Hubbard $U$ ($g=K=0$),
it is found that the metal-insulator transition in A$_3$C$_{60}$
takes place at the upper range of what is believed to be physical
values of $U/W$, while for A$_4$C$_{60}$ this happens at the 
lower range of these parameters. This agrees nicely with 
the fact that A$_3$C$_{60}$ is a metal but A$_4$C$_{60}$ is
an insulator. However, to explain why A$_4$C$_{60}$ is not 
antiferromagnetic, we need to include the coupling to the
Jahn-Teller H$_g$ phonons. This substantially lowers the
critical $U_c$ for both systems, and it puts $U_c/W$
of A$_3$C$_{60}$ at the lower range of the physical parameters.
This puts our understanding of A$_3$C$_{60}$ being a metal
into question. We find, however, that the competition 
between the Jahn-Teller effect and the Hund's rule coupling 
increases $U_c$ again. The coupling to the $t_{1u}$ plasmons
in A$_3$C$_{60}$ should lead to an additional increase of 
$U_c$ in this system. This makes it understandable that 
A$_3$C$_{60}$ can be a metal.

\section{Acknowledgements}
This work has been supported by the Max-Planck-Forschungspreis.

\begin{table}
\caption[]{The ground-state energy $E(N)$ for $N$ electrons in the
$E\times E$ case. The quantity $N(N-1)U_{av}(2)/2$ has been subtracted, 
where $U_{av}(2)=U_{xy}+K/3$ is the average interaction for a full shell.}
\begin{tabular}{ccc}
\multicolumn{1}{c}{$N$} &
\multicolumn{2}{c}{$\tilde E(N)\equiv E(N)-N(N-1)U_{av}(2)/2$}  \\
    &  \rm Low spin   &  High spin   \\
\tableline
   1    & $-2E_{JT}$  &               \\
   2    & $-8E_{JT}+{8\over 3}K     $  &  $ -{4\over 3}K$  \\
   3    & $-2E_{JT} $     &          \\
\end{tabular}\label{table1}
\end{table}

\begin{table}
\caption[]{The ground-state energy $E(N)$ for $N$ electrons in the
$T\times H$ case. The quantity $N(N-1)U_{av}(3)/2$ has been subtracted,
where $U_{av}(3)=U_{xy}$ is the average interaction for a full shell.}
\begin{tabular}{ccc}
\multicolumn{1}{c}{$N$} &
\multicolumn{2}{c}{$\tilde E(N)\equiv E(N)-N(N-1)U_{av}(3)/2$}  \\
    &  \rm Low spin   &  High spin   \\
\tableline
   1    & $-{5\over 2}E_{JT}$  &               \\
   2    & $-10E_{JT}+4K     $  &  $ -{5\over 2}E_{JT}-K$  \\
   3    & $-{15\over 2}E_{JT}+2K$ &$ -3K$                 \\
   4    & $-10E_{JT}+4K     $  &  $ -{5\over 2}E_{JT}-K$  \\
   5    & $-{5\over 2}E_{JT}$  &               \\
\end{tabular}\label{table2}
\end{table}
\begin{table}
\caption[]{$E_g(n)-U_{av}(2)-d_2(n)t$ for the $A\times E$ and the $E\times E$ 
two-site models as a function of the filling $n$, where $E_g$ is the 
band gap,  $U_{av}(2)$ is the average Coulomb interaction and $d_2(n)t$ 
is the hopping contribution, with $d_2(n)=$ -3 and -4 for $n=1$ and 2,
respectively. The results are symmetric around half-filling $n=2$.}
\begin{tabular}{cccc}
\multicolumn{1}{c}{Syst} & \multicolumn{1}{c}{$n$}& \multicolumn{2}{c}
{$E_g(n)-U_{av}(2)-d_2(n)$} \\
\tableline
  &  & $K\le 2E_{JT}$  &  $K>2E_{JT}$   \\
\tableline
  $A\times E$ & 1& \multicolumn{2}{c}{$-E_{JT}$}\\
  $A\times E$ & 2& \multicolumn{2}{c}{$-E_{JT}+{16\over 3}K$}\\
  $E\times E$ & 1& \multicolumn{2}{c}{$ 6E_{JT}$}   \\
  $E\times E$ & 2& $ 30E_{JT}-{32\over 3}K$ & $2E_{JT}+
{16\over 3}K$ \\ 
\end{tabular}\label{table3}
\end{table}
\begin{table}
\caption[]{$E_g(n)-U_{av}(3)-d_3(n)t$ for the $T\times H$ two-site 
model as a function of the filling $n$. The hopping contribution to the 
gap is given by $d(n)t$, where $d_3(n)=$ -3, -5 and -6 for $n=$ 1, 2 and 3,
respectively. The results are symmetric around half-filling $n=3$.}
\begin{tabular}{cccc}
\multicolumn{1}{c}{} & \multicolumn{3}{c}{$E_g(n)-U_{av}(3)-d_3(n)t$} \\
\tableline
$n$ & $K\le {3\over 2}E_{JT}$ & ${3\over 2}E_{JT}< K \le {9\over 4}E_{JT}$
& $K>{9\over 4}E_{JT}$ \\
\tableline
1 & \multicolumn{2}{c}{$5E_{JT}+{2\over 3}K$} & ${35\over 4}E_{JT}- K$ \\
2 & $35E_{JT}-{46\over 3}K$ & $ 5E_{JT}+
{14\over 3}K $ & $ {35\over 4}E_{JT}+3K$ \\
3 & ${55\over 2}E_{JT}-8K$ & \multicolumn{2}{c}{
$-{5\over 2}E_{JT}+12 K$}\\ 
\end{tabular}\label{table4}
\end{table} 
\end{multicols}
\end{document}